\documentclass[11pt]{article}
\usepackage[colorlinks=true,citecolor=blue,linkcolor=blue]{hyperref}
\usepackage{multirow}
\usepackage[left=2cm,right=2cm,top=2cm,bottom=2cm]{geometry}
\usepackage{cite}
\usepackage{psfrag}
\usepackage[demo]{graphicx}
\usepackage{graphicx}
\usepackage{graphics}
\usepackage{epsfig}
\usepackage{booktabs}
\usepackage{amsmath,amsfonts,amssymb}
\usepackage{pstcol,pst-fill,pst-grad}
\usepackage{pstricks,pst-fill,pst-grad}
\usepackage{euscript}
\usepackage{pstricks}
\usepackage{wrapfig}
\usepackage{slashed}
\usepackage[toc,page]{appendix}
\usepackage{here}

\begin{document}
	\title{\vspace{-3cm}
		\hfill\parbox{4cm}{\normalsize \emph{}}\\
		\vspace{1cm}
		{  Higgs-strahlung boson production in the presence of a circularly polarized laser field}}
	\vspace{2cm}
	
	\author{ M. Ouhammou,$^1$ M. Ouali,$^1$  S. Taj,$^1$ and B. Manaut$^{1,}$\thanks{Corresponding author, E-mail: b.manaut@usms.ma} \\
		{\it {\small$^1$ Sultan Moulay Slimane University, Polydisciplinary Faculty,}}\\
		{\it {\small Research Team in Theoretical Physics and Materials (RTTPM), Beni Mellal, 23000, Morocco.}}\\				
	}
	\maketitle \setcounter{page}{1}
\date{}
\begin{abstract}
In the framework of the electroweak standard model, we investigated, in the center of mass frame, the Higgs boson production in the presence of an intense laser field via ${e}^{+} {e}^{-}$  annihilation $({e}^{+} {e}^{-}\rightarrow ZH)$. By comparing our results with those obtained by Djouadi \cite{1} for laser-free process, we show that the circularly polarized laser field affects significantly the $s$-channel Higgs boson production. We find that for a given number of exchanged photons, laser field strength and frequency, the total cross section decreases by several order of magnitude. These effects of laser field on cross section are found to be consistent with what was found for muon pair production via QED process in the presence of a circularly polarized laser field \cite{2}.
\end{abstract}
\maketitle
\section{Introduction}
\par The effect of the laser on the behavior of physical systems is relevant in the study of phenomena related to the electroweak standard model of particle physics, where the electromagnetic field could have an important impact on particles properties and its interactions. It is well known that the electromagnetic field affects the physical properties of particles, then it would be interesting to investigate how the transition probability is modified in the presence of an ultraintense laser fields. This could eventually play an important role especially on high energy  production and decay processes. These effect on particle physics interaction has been a subject of several works. In\cite{3}, the authors showed that the insertion of an external field has a big impact on $Z$ boson decay as it enhances its mode decay and affects its life time. The same results was obtained for muon decay \cite{4} and pion decay \cite{5}. The muon pair production via electron-positron annihilation in QED was studied in the presence of both circularly \cite{2} and linearly \cite{6} polarized laser field. It is found that the insertion of a linearly polarized laser field enhances the total cross section of the muon pair production while the circularly polarized laser field reduces the total cross section by several order of magnitude. In this respect, we have investigated the electroweak process ${e}^{+} {e}^{-}\rightarrow ZH$ in the presence of a circularly polarized laser field to show how the cross section of the Higgs production is affected.

\par Higgs boson, which play a key role in the symmetry breaking mechanism by generating particle masses in the standard model, was discovered by ATLAS and CMS experiment in 2012 \cite{7,8} with a mass around 125 GeV. The noble prize in physics was awarded to Peter Higgs and Francois Englert in 2013 for their works in identifying and discovering it. Many of Higgs boson properties have been measured by ATLAS and CMS collaboration \cite{9,10,11,12,13,14,15,16,17} using data collected during run 1 and run 2 of the Large Hadron Collider (LHC); moreover, all these properties are found to be consistent with the prediction of the standard model of particle physics. As a result, the standard model  get the full recognition as a very successful model even though there are hints of existence of new physics beyond the standard model. The latter is a quantum field theory that unifies three of the four fundamental interactions known today, which are the weak nuclear force, the electromagnetic force and the strong fore. It gives an almost complete description of the elementary particle physics. The Large Electron-Positron Collider (LEP), which operated from 1989 to 2000, was the first particle accelerator to have significant reach into the potential mass range of the Higgs boson. It makes significant headway in the search, determinating that the mass should be larger than 114 GeV. The future Circular Electron Positron collider (CEPC) is mainly designed to operate at center of mass energies around $240-250\,GeV$ during 10 years in three different modes which are ${e}^{+} {e}^{-} \rightarrow {Z} {H}$, ${e}^{+} {e}^{-} \rightarrow {Z}$ and ${e}^{+} {e}^{-} \rightarrow {W}^{+} {W}^{-}$. The first 7 years will dedicated to the production of more than one million Higgs bosons via the Higgs strahlung process. This fact highly motivate us to introduce the circularly polarized laser field in Higgs strahlung process ${e}^{+} {e}^{-} \rightarrow {Z}{H}$ in order to show its effect on the Higgs production cross section.

\par The aim of this paper is to study the Higgs boson production via the Higgs strahlung process in the framework of electroweak standard model in the presence of a circularly polarized laser field. The Higgs boson is produced when an electron and positron annihilate to create a Higgs boson and a supermassive $Z$ boson. This process is the main mechanism by which a Higgs boson can be produced in the future Circular Electron Positron collider (CEPC). Even-though the circularly polarized laser field obviously decreases the cross section, the results found may help in particle physics experiments, especially at CEPC where the  the first run will dedicated to the Higgs strahlung production. We hope that we study this process in the presence of a linearly polarized field in a future work to show its effect on Higgs production cross section.

\par The remainder of this paper is presented in three sections: In section 2, we begin by presenting the analytical calculation of the total cross section of the process ${e}^{+} {e}^{-} \rightarrow {Z} {H}$ in the presence of a circularly polarized laser field. Then, the data obtained are presented and discussed in section 3. Finally, a brief conclusion is given in section 4. Some Bessel functions multiplying coefficients are listed in the appendix. In this work the natural units are adopted ($\hbar = c = 1$). The choice made for Livi-Civita tensor is that $\epsilon^{0123}=1$ and the metric $g^{\mu\nu}$ is chosen such that $g^{\mu\nu}=(1,-1,-1,-1)$.
\section{Outline of the theory}\label{sec:theory}
The cross section is one of the most important measured quantities in production scattering processes. In  this research paper's part, we present the analytical calculation of the total cross section of Higgs strahlung process in the presence of a circularly polarized laser field. We find that this quantity depends on the energy of the incident particles, i.e., the center of mass energy and the laser's parameters such as the exchanged photons number, the laser field strength and the laser frequency.
\subsection*{A-Description of the laser field and wave functions}
\begin{figure}[H]
  \centering
      \includegraphics[width=0.5\textwidth]{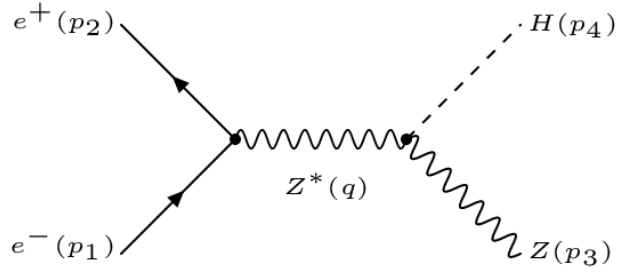}
  \caption{Feynman diagram  for $s$-channel Higgs production in the standard model  during the $e^{+}e^{-}$ collision (annihilation) in the lowest order.}
\end{figure}
$p_{1}$, $p_{2}$, $p_{3}$ and $p_{4}$ are the free four vector momenta of the electron, positron, $Z$-boson and Higgs boson, respectively. In the present work, we consider an electromagnetic, circularly polarized and monochromatic wave field propagating along the $z$-axis. It is represented by the classical four-vector potential $A^{\mu}$ such that:
\begin{equation}
A^{\mu}(\phi)=a_{1}^{\mu}\cos\phi+a_{2}^{\mu}\sin\phi \hspace*{1cm};\hspace*{1cm} \phi=(k.x)
\label{1}
\end{equation}
 Where $k^{\mu}$ is taken to be parallel to the $ z$-axis $(k=(\omega,0,0,\omega))$ and $(k^{2}=0)$,  $\phi$ is the phase of the laser field and $\omega$ its frequency. The Lorentz gauge is satisfied $\partial_{\mu}A^{\mu}=0$, which implies that $(k_{\mu}A_{\mu}=0;\, a_{1}.k=0;\, a_{2}.k=0)$. $a_{1,2}^{\mu}$ are the polarization four vectors chosen as  $a_{1}^{\mu}=(0,a,0,0)$ and $a_{2}^{\mu}=(0,0,a,0)$ where $a$ is the amplitude of the four-vector potential. They are orthogonals, which implies that $(a_{1}.a_{2})=0$ and $a_{1}^{2}=a_{2}^{2}=a^{2}=-|\mathbf{a}|^{2}=-\big(\frac{\epsilon_{0}^{2}}{\omega}\big)$ with $\epsilon_{0}$ is the laser's field strength.\\
 The incoming electron and positron are electrically charged particle and its antiparticle, respectively, with mass $m_{e}$. They are described by Dirac-Volkov states \cite{18} which satisfies the Dirac equation in an external field
 \begin{equation}
 \Big[(p_{i}-eA)^{2}-m_{e}^{2}-\dfrac{ie}{2}F_{\mu\nu}\sigma^{\mu\nu} \Big]\psi_{p_{i},s_{i}}(x)=0
 \label{2}
 \end{equation}
Here, the indices $p_{i}(i=1, 2)$ and $ s_{i}(i=1, 2)$ refer to the particle momentum and its spin outside the electromagnetic field. Dressed by such a laser field, the incoming electron and positron can be viewed as Dirac-Volkov states \cite{18} described by:
\begin{equation}
\begin{cases}
\psi_{p_{1},s_{1}}(x)= \Big[1-\dfrac{e \slashed k \slashed A}{2(k.p_{1})}\Big] \frac{u(p_{1},s_{1})}{\sqrt{2Q_{1}V}} \exp^{iS(q_{1},s_{1})}&\\
\psi_{p_{2},s_{2}}(x)= \Big[1+\dfrac{e \slashed k \slashed A}{2(k.p_{2})}\Big] \frac{v(p_{2},s_{2})}{\sqrt{2Q_{2}V}} \exp^{iS(q_{2},s_{2})}
\end{cases}
\label{3}
\end{equation}
where the quantities $S(q_{i},s_{i})(i=1,2)$ are defined as:
\begin{equation}
\begin{cases}
S(q_{1},s_{1})=- q_{1}x +\frac{e(a_{1}.p_{1})}{k.p_{1}}\sin\phi - \frac{e(a_{2}.p_{1})}{k.p_{1}}\cos\phi &\\
S(q_{2},s_{2})=+ q_{2}x +\frac{e(a_{1}.p_{2})}{k.p_{2}}\sin\phi - \frac{e(a_{2}.p_{2})}{k.p_{2}}\cos\phi
\end{cases}
\label{4}
\end{equation}
 $u(p_{1},s_{1})$ and $v(p_{2},s_{2})$ represent the bispinor of the electron and positron, respectively, with a momentum $p_{i}(i=1,2)$ and spin $s_{i}(i=1,2)$. They satisfy the following equation: $\sum_{s}^{}u(p_{1},s_{1})\bar{u}(p_{1},s_{1})=(\slashed p_{1}-m_{e})$ and $\sum_{s}^{}v(p_{2},s_{2})\bar{v}(p_{2},s_{2})=(\slashed p_{2}+m_{e})$\\
The four vector $q_{i}=(Q_{i},q_{i})(i=1,2)$  is the effective four vector momentum of the electron and the positron inside laser field with:
\begin{equation}
q_{i}=p_{i}+\dfrac{e^{2}a^{2}}{2(k.p_{i})}k
\label{5}
\end{equation}
Its zero component $q_{i}^{0}=Q_{i}$ is the effective energy appearing in the normalization factor of the Volkov wave function. The effective mass can be obtained as:
\begin{equation}
 m_{e}^{*^{2}}= q_{i}^{2}=(m_{e}^{2}+e^{2}a^{2})
 \label{6}
\end{equation}
The scattered Higgs boson $H$ and $Z^{0}$-boson are massive particles with spin $0$ and $1$, respectively. They are electrically neutral which means that they will never interact with the laser field and they are described by the following equations:
\begin{equation}
Z^{\nu}(y)=\dfrac{\epsilon^{\nu}(p_{3},\lambda)}{\sqrt{2 Q_Z V}} e^{-ip_{3}y} \hspace*{0.5cm};\hspace*{0.5cm}  \\\ \phi(y)=\dfrac{1}{\sqrt{2 Q_H V}} e^{-ip_{4}y}
\label{7}
\end{equation}
Where $p_{3}$ and $p_{4}$ are the four vector momenta of the  $Z^{0}$-boson and Higgs boson, respectively. $Q_{Z}=p_{3}^{0}$ and $Q_{H}=p_{4}^{0}$. $\epsilon^{\nu}(p_{3},\lambda)$ is the  $Z^{0}$-boson polarisation vector and  $(\lambda)$ denotes its  polarization.
\subsection*{B- Scattering amplitude and cross section}
The lowest scattering-matix element \cite{19} for  the laser assisted Higgs strahlung process can be written as:

\begin{equation}
S_{fi}({e}^{+}{e}^{-}\rightarrow ZH)=\frac{g M_{Z}^{2}}{2\cos_{\theta_W} v} \int_{}^{} d^4x \int d^4y \phi^{*}(y) Z^{* \nu}  (y)D_{\mu \nu}(x-y) \bar{\psi}_{p_{2},s_{2}}(x) \Big[ \gamma^{\nu} (g_v^{e} -g_a^{e}\gamma^{5}) \Big] \psi_{p_{1},s_{1}}(x)
\label{8}
\end{equation}

where $M_{Z}$ is the  scattered   $Z$-boson mass, {$ v=(\sqrt{2}G_{F})^{-\frac{1}{2}} $} is the vacuum expectation value, {$ \theta_{W} $} is the Weinberg angle, $ g_{v}^{e}=-1+4 \sin^{2}\theta_{w} $ and $ g_{a}^{e}=1 $ are the vector and axial vector coupling constants, respectively.  {$ g $} is the electroweak coupling constant which is related to the electroweak mixing angle by $g^{2}=\frac{e^{2}}{\sin^{2}\theta_{w}}=\frac{8G_{F}M_{Z}^{2}\cos^{2}_{\theta_{w}}}{\sqrt{2}}$ where $e=1eV$ and $G_{F}=1.166 3787 \times 10^{-5} GeV^{-2}$ are the electron's charge and the Fermi coupling constant, respectively.

 $ D_{\mu\nu}(x-y) $ is the  $ Z^{*} $-boson propagator  \cite{19}. It is given by:
\
\begin{equation}
D_{\mu\nu}(x-y)=\int \dfrac{d^{4}q}{(2\pi)^4} \frac{e^{-iq(x-y)}}{q^{2}-M_{Z}^{2}}\Bigg[-ig_{\mu\nu}+i\dfrac{q_{\mu}q_{\nu}}{M_{Z}^{2}}\Bigg]
\label{9}
\end{equation}

where $q$ is the four vector momentum of the $ Z^{*} $-boson propagator.\\

 By inserting the equations (\ref{3}), (\ref{7}) and (\ref{9}) into the scattering-matrix element given by equation (\ref{8}), it becomes:
\small
\begin{equation}
S_{fi}({e}^{+}{e}^{-}\rightarrow HZ)=\frac{g M_{Z}^2}{2\cos_{\theta_W} v} \dfrac{1}{\sqrt{16V^{4}Q_{Z}Q_{H}Q_{1}Q_{2}}}\Bigg[ \dfrac{1}{(q_{1}+q_{2})^{2}-M_{Z}^{2}}  \Bigg]  M_{fi}^{n} \, (2\pi)^{4}\, \\ \delta^{4}\big( p_{3}+p_{4}-q_{1}-q_{2}-nk \big)
\label{10}
\end{equation}
\normalsize
Where $(n)$ is the number of photons that are exchanged with the charged incident particles, with $(n)$ might be either positive or negative. $n<0$ refers to the emission of $n$ photons and $n>0$ refers to the absorption.
 $ M_{fi}^{n} $ is the amplitude of the Higgs strahlung process which can be decomposed in terms of Bessel function series and it is given by:
\begin{equation}
M_{fi}^{n} =\epsilon ^{*\nu}(p_{3},\lambda)\bar{v}(p_{2},s_{2})\Lambda^{n}_{\mu}u(p_{1},s_{1})
\label{11}
\end{equation}
with $\Lambda^{n}_{\mu}$ is derived by using the generating function for the Bessel functions \cite{20}
\begin{equation}
\sum_{n=-\infty}^{+\infty}J_{n}(z)e^{-in\phi}=e^{-iz\sin\phi}
\label{12}
\end{equation}
We find that:
\begin{equation}
 \Lambda_{\mu}^{n}=C^{0}_{\mu}\,\xi_{0n}(z)+C^{1}_{\mu}\,\xi_{1n}(z)+C^{2}_{\mu}\,\xi_{2n}(z)
 \label{13}
 \end{equation}
  Where the quantities  $C^{0}_{\mu}$, $C^{1}_{\mu}$   and   $C^{2}_{\mu}$ are explicitly expressed by:\\
\begin{equation}
\begin{cases}C^{0}_{\mu}=\gamma_{\mu}(g_{v}^{e}-g_{a}^{e}\gamma^{5})+2c_{p_{1}}c_{p_{2}}a^{2}k_{\mu}\slashed k(g_{v}^{e}-g_{a}^{e}\gamma^{5})   &\\C^{1}_{\mu}=c_{p_{1}}\gamma_{\mu}(g_{v}^{e}-g_{a}^{e}\gamma^{5})\slashed k\slashed a_{1}-c_{p_{2}}\slashed a_{1}\slashed k \gamma_{\mu}(g_{v}^{e}-g_{a}^{e}\gamma^{5})   &\\C^{2}_{\mu}=c_{p_{1}}\gamma_{\mu}(g_{v}^{e}-g_{a}^{e}\gamma^{5})\slashed k\slashed a_{2}-c_{p_{2}}\slashed a_{2}\slashed k \gamma_{\mu}(g_{v}^{e}-g_{a}^{e}\gamma^{5}) \end{cases}
\label{14}
\end{equation}
With $ c_{p_{1}}=\dfrac{e}{2(k.p_{1})} $  and $ c_{p_{2}}=\dfrac{e}{2(k.p_{2})} $ \\
The quantities $\xi_{0n}(z)$, $\xi_{1n}(z)$ and $\xi_{2n}(z)$ are expressed in terms of Bessel functions by:
\small
\begin{equation}
\left.
  \begin{cases}
     \xi_{0n}(z) \\
      \xi_{1n}(z) \\
      \xi_{2n}(z)
  \end{cases}
  \right\} = \left.
  \begin{cases}
     J_{n}(z)e^{-in\phi _{0}}\\
    \frac{1}{2}\Big(J_{n+1}(z)e^{-i(n+1)\phi _{0}}+J_{n-1}(z)e^{-i(n-1)\phi _{0}}\Big) \\
     \frac{1}{2\, i}\Big(J_{n+1}(z)e^{-i(n+1)\phi _{0}}-J_{n-1}(z)e^{-i(n-1)\phi _{0}}\Big)
  \end{cases}
  \right\}
  \label{15}
\end{equation}
\normalsize
The argument of the Bessel function is expressed by:
$z=\sqrt{\alpha_{1}^{2}+\alpha_{2}^{2}}$ and $\phi_{0}= \arctan(\frac{\alpha_{2}}{\alpha_{1}})$, where:
\begin{center}
$\alpha_{1}=\dfrac{e(a_{1}.p_{1})}{(k.p_{1})}-\dfrac{e(a_{1}.p_{2})}{(k.p_{2})}$ \qquad \qquad $\alpha_{2}=\dfrac{e(a_{2}.p_{1})}{(k.p_{1})}-\dfrac{e(a_{2}.p_{2})}{(k.p_{2})}$\\
\end{center}
In the center of mass frame, the differential cross section is obtained by dividing the square of  the S-matrix element by $ VT $ to obtain the transition probability per volume, by $|J_{inc}|=(\sqrt{(q_{1}q_{2})^{2}-m_{e}^{*^{4}}}/{Q_{1}Q_{2}V})$, and by the particle density $\rho=V^{-1}$ and finally one has to average over the initial spins and sum over the final ones. We obtain:
\small
\begin{equation}
\dfrac{d\sigma_{n}}{d\Omega}=\dfrac{g^{2}M_{Z}^{4}}{64\cos_{\theta_{W}}^{2}v^{2}}\Bigg(  \dfrac{1}{(q_{1}+q_{2})^{2}-M_{Z}^{2}} \Bigg)^{2}  \dfrac{1}{\sqrt{(q_{1}q_{2})^{2}-m_{e}^{*^{4}}}} \big|\overline{M_{fi}^{n}} \big|^{2} \int \dfrac{2|\mathbf{p}_{4}|^{2}d|\mathbf{p}_{4}|}{(2\pi)^{4}Q_{H}} \int \dfrac{d^{3}p_{3}}{Q_{Z}}\\ \times\delta^{4}(p_{3}+p_{4}-q_{1}-q_{2}-nk)
\label{16}
\end{equation}
\normalsize
By integrating over $d^{3} p_{3}$, the differential cross section becomes:
\begin{equation}
\dfrac{d\sigma_{n}}{d\Omega}=\dfrac{g^{2}M_{Z}^{4}}{64\cos_{\theta_{W}}^{2}v^{2}}\Bigg(  \dfrac{1}{(q_{1}+q_{2})^{2}-M_{Z}^{2}} \Bigg)^{2}  \dfrac{1}{\sqrt{(q_{1}q_{2})^{2}-m_{e}^{*^{4}}}} \big|\overline{M_{fi}^{n}} \big|^{2}  \int \dfrac{2|\mathbf{p}_{4}|^{2}d|\mathbf{p}_{4}|}{(2\pi)^{2}Q_{H}} \delta(Q_{Z}+Q_{H}-Q_{1}-Q_{2}-n\omega)
\label{17}
\end{equation}
With the argument of the delta function in the formula (\ref{16}) determines the law of the momentum conservation:
\begin{equation}
\mathbf{p_{3}}+\mathbf{p_{4}}-\mathbf{q_{1}}-\mathbf{q_{2}}-\mathbf{nk}=0
\label{18}
\end{equation}
The remaining integral in the differential cross section expression can be performed by using the well known formula \cite{19} given by:
\begin{equation}
 \int d\mathbf x f(\mathbf x) \delta(g(\mathbf x))=\dfrac{f(\mathbf x)}{|g^{'}(\mathbf x)|_{g(\mathbf x)=0}}
 \label{19}
\end{equation}
thus, we get:
\begin{equation}
\dfrac{d\sigma_{n}}{d\Omega}=\dfrac{g^{2}M_{Z}^{4}}{64\cos_{\theta_{W}}^{2}v^{2}}\Bigg(  \dfrac{1}{(q_{1}+q_{2})^{2}-M_{Z}^{2}} \Bigg)^{2}  \dfrac{1}{\sqrt{(q_{1}q_{2})^{2}-m_{e}^{*^{4}}}} \big|\overline{M_{fi}^{n}} \big|^{2}  \dfrac{2|\mathbf{p_{4}}|^{2}}{(2\pi)^{2}Q_{H}}\dfrac{1}{\big|g^{'}(|\mathbf{p}_{4}|)\big|_{g(|\mathbf{p}_{4}|)=0}}
\label{20}
\end{equation}
with
\begin{equation}
 g^{'}(|\mathbf{p}_{4}|)=\dfrac{-4 e^{2}a^{2}}{\sqrt{s}}\dfrac{|\mathbf{p}_{4}|}{\sqrt{|\mathbf{p}_{4}|^{2}+M_{H}^{2}}}-\dfrac{2|\mathbf{p}_{4}|(\sqrt{s}+n\omega)}{\sqrt{|\mathbf{p}_{4}|^{2}+M_{H}^{2}}}
 \label{21}
\end{equation}
where $( \sqrt{s}) $ is the center of mass energy.\\
The equation (\ref{20}) indicates that the total differential cross section can be decomposed into series of partial discrete differential cross section for different number of exchanged photons.\\
The total cross section is obtained by performing a numerical integration over the solid angle $d\Omega=\sin\theta d\theta d\phi$. The term  $\big|\overline{M_{fi}^{n}} \big|^{2}$ appearing in the cross section can be evaluated as follows:
\begin{equation}
\big|\overline{M_{fi}^{n}} \big|^{2}=\dfrac{1}{4}\sum_{n=-\infty}^{+\infty}\sum_{\lambda}\sum_{s}\big|M_{fi}^{n} \big|^{2}=\frac{1}{4}\sum_{n=-\infty}^{+\infty}\big(  -g^{\mu\nu}+\dfrac{p_{3}^{\mu}p_{3}^{\nu}}{M_{Z}^{2}}  \big)Tr\big[ (\slashed p_{1}-m_{e}) \Lambda^{n}_{\mu}(\slashed p_{2}+m_{e})\bar{\Lambda}^{n}_{\nu}\big]
\label{22}
\end{equation}
Where, $\bar{\Lambda}_{\nu}^{n}=\bar{C^{0}_{\nu}}\,\xi_{0n}^{*}(z)+\bar{C^{1}_{\nu}}\,\xi_{1n}^{*}(z)+\bar{C^{2}_{\nu}}\,\xi_{2n}^{*}(z)$\\
and :
\small
\begin{equation}
\begin{cases}\bar{C^{0}_{\nu}}=\gamma_{\nu}(g_{v}^{e}-g_{a}^{e}\gamma^{5})+2c_{p_{1}}c_{p_{2}}a^{2}k_{\nu}\slashed k(g_{v}^{e}-g_{a}^{e}\gamma^{5})   &\\
\bar{C^{1}_{\nu}}=c_{p_{1}} \slashed a_{1} \slashed k \gamma_{\nu}(g_{v}^{e}-g_{a}^{e}\gamma^{5}) -c_{p_{2}}\gamma_{\nu}(g_{v}^{e}-g_{a}^{e}\gamma^{5})\slashed k\slashed a_{1}   &\\
\bar{C^{2}_{\nu}}=c_{p_{1}} \slashed a_{2} \slashed k \gamma_{\nu}(g_{v}^{e}-g_{a}^{e}\gamma^{5}) -c_{p_{2}}\gamma_{\nu}(g_{v}^{e}-g_{a}^{e}\gamma^{5})\slashed k\slashed a_{2} \end{cases}
\label{23}
\end{equation}
\normalsize
 Because of its hardness, the trace calculation is performed by using FeynCalc program. The result obtained is too long to be presented in this paper, but, in general, it has the following form:
\begin{equation}
\big|\overline{M_{fi}^{n}} \big|^{2}=\frac{1}{4}\sum_{n=-\infty}^{+\infty}\Big[ AJ_{n}^{2}(z)+BJ_{n+1}^{2}(z)+CJ_{n-1}^{2}(z)+DJ_{n}(z)J_{n+1}(z)+EJ_{n}(z)J_{n-1}(z)+FJ_{n-1}(z)J_{n+1}(z) \Big]
\label{24}
\end{equation}
The coefficients A, B, C, D, E and F are calculated using FeynCalc tools. We give here the expression of the first coefficient multiplied by $J_{n}^{2}(z)$, the others are given in the appendix.
\begin{eqnarray}
A&=&\nonumber\dfrac{2}{((k.p_{1}) (k.p_{2}) M_{Z}^{2})}\Big[(a^{4} e^{4} (g_{a}^{e^{2}} + g_{v}^{e^{2}}) (k.p_{3})^2 +2 (k.p_{1}) (k.p_{2}) (M_{Z}^{2} (g_{a}^{e^{2}} (-3 m_{e}^{2} + (p_{1}.p_{2})) \\&+& g_{v}^{e^{2}} (3 m_{e}^{2}+(p_{1}.p_{2})))+2 (g_{a}^{e^{2}} + g_{v}^{e^{2}}) (p_{1}.p_{3}) (p_{2}.p_{3}))+2 a^{2} e^2 (g_{a}^{e^{2}} (-2 (k.p_{1}) (k.p_{2}) M_{Z}^{2} \\&+&\nonumber(k.p_{3}) ((k.p_{3}) m_{e}^{2} - (k.p_{3}) (p_{1}.p_{2}) + (k.p_{2}) (p_{1}.p_{3}) + (k.p_{1}) (p_{2}.p_{3})))+g_{v}^{e^{2}} (-2 (k.p_{1}) (k.p_{2})\\&\times &\nonumber M_{Z}^{2}(k.p_{3}) (-(k.p_{3}) (m_{e}^{2} + (p_{1}.p_{2})) + (k.p_{2}) (p_{1}.p_{3})+ (k.p_{1}) (p_{2}.p_{3})))))\Big]
\label{25}
\end{eqnarray}
We have checked the validity of the sum-rule \cite{21} for multiphotonic Higgs strahlung process, we find after calculation that:
\begin{equation}
\sum_{n=-\infty}^{+\infty}\dfrac{d\sigma_{n}}{d\Omega}=\Big(\dfrac{d\sigma}{d\Omega}\Big)_{laser-free}
\label{26}
\end{equation}
and
\begin{equation}
\sum_{n=-\infty}^{+\infty}\sigma_{n}=\big(\sigma \big)_{laser-free}
\label{(27)}
\end{equation}
\section{Results and Discussion}\label{sec:results}
\begin{figure}[H]
  \centering
      \includegraphics[scale=0.8]{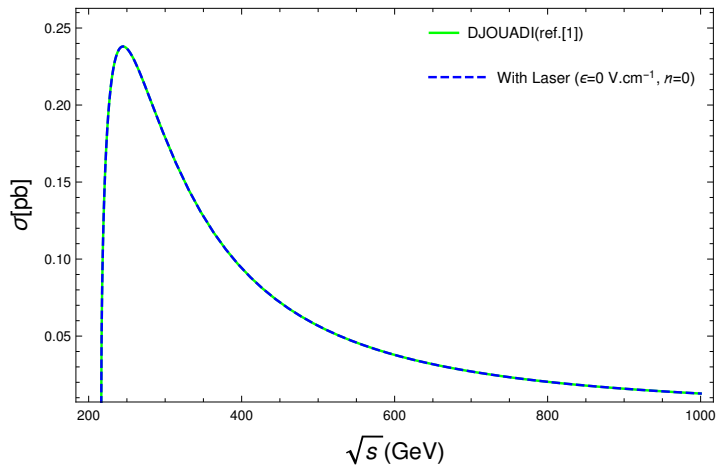}
  \caption{Comparaison between DJOUADI's total cross section \cite{1} and the total cross section in the presence of the laser field  for the Higgs strahlung process  $e^{+}e^{-}\rightarrow ZH$ as a function of the center of mass energy  $\sqrt{s}$. The strength of the laser field and the number of exchanged photons are  $\epsilon_{0}=0\,\,V.cm^{-1}$ and $n=0$, respectively.}
  \label{fig2}
\end{figure}

In this section, we discuss the numerical results obtained for one of the most fundamental process in the standard model in the presence of a circularly polarized laser field. In the center of mass frame, we show the dependence of the total cross section on the photon number transfer, the number of exchanged photons, the laser field strength and its frequency. The numerical data and curves presented in this section are obtained by using Mathematica. All curves are plotted by taking the center of mass energy as $\sqrt{s}=250\,GeV$ except in Fig.\ref{fig2} and Fig.\ref{fig6}, eventhough we can use various other energy point such as $240GeV$ to satisfy the CEPC concerns.
Figure \ref{fig2} shows the total cross section for the Higgs strahlung production in the presence of a circularly polarized laser field  in comparison with that obtained by DJOUADI in \cite{1}. The laser's parameters are chosen as follows: the laser strength ($\epsilon_{0}=0\,V.cm^{-1}$) and the number of exchanged photons is ($n=0$). The aim of this comparison is to check that our results converge to those obtained in the absence of the laser field by taking  $\epsilon_{0}=0\,\,V.cm^{-1}$ and $n=0$. According to the Fig.\ref{fig2}, the two curves are in full agreement for every center of mass energy  $\sqrt{s}$. This comparison makes as more comfortable not only about our analytical calculation but also about the results that will be discussed bellow.
\begin{figure}[H]
  \centering
      \includegraphics[scale=0.8]{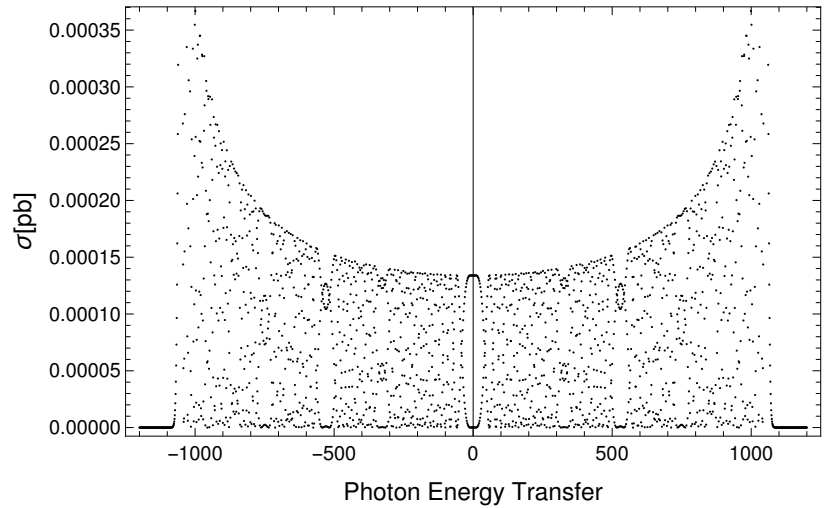}
  \caption{The total cross section of the Higgs strahlung process  $e^{+}e^{-}\rightarrow ZH$ in natural units as a function of the photons energy transfer number for the center of mass energy of $(\sqrt{s}=250GeV)$. The laser field strength and the laser frequency are chosen as $\epsilon_{0}=10^{7}V.cm^{-1}$ and $\omega=1.17\, eV$, respectively.}
  \label{fig3}
  \end{figure}
  \begin{figure}[H]
  \centering
      \includegraphics[scale=0.8]{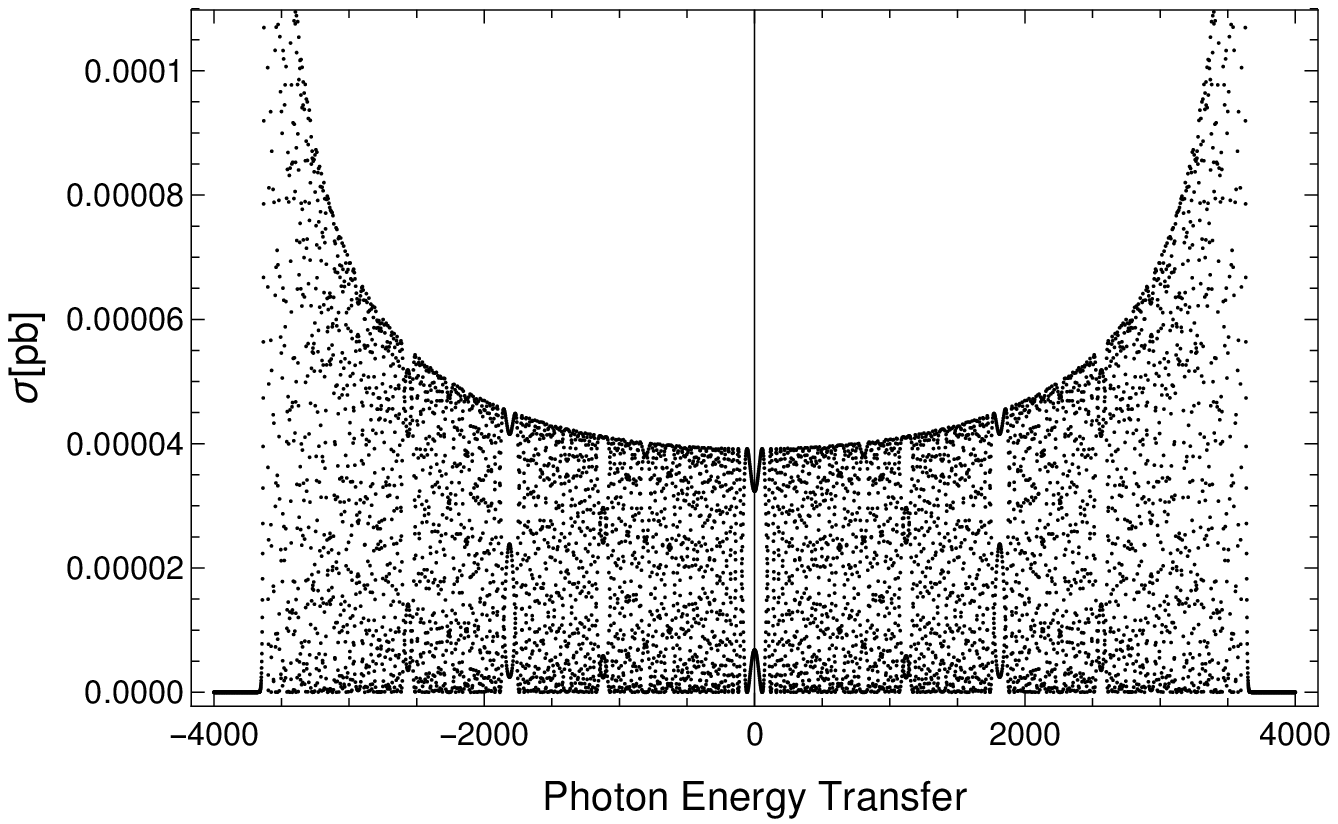}
  \caption{The total cross section of the Higgs strahlung process  $e^{+}e^{-}\rightarrow ZH$ in natural units as a function of the photons energy transfer number for the center of mass energy of $(\sqrt{s}=250GeV)$. The laser field strength and the laser frequency are chosen as $\epsilon_{0}=10^{8}V.cm^{-1}$ and $\omega=2\, eV$ , respectively.}
  \label{fig4}
  \end{figure}
  \begin{figure}[H]
  \centering
      \includegraphics[scale=0.8]{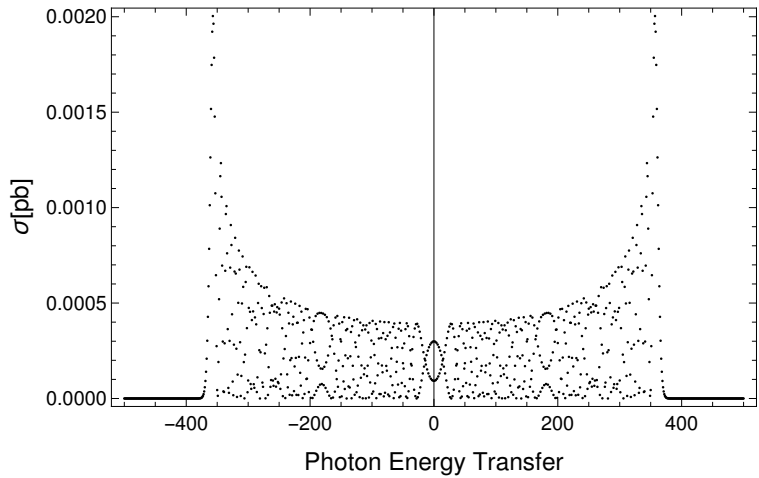}
  \caption{The total cross section of the Higgs strahlung process  $e^{+}e^{-}\rightarrow ZH$ in natural units as a function of the photons energy transfer number for the center of mass energy of $(\sqrt{s}=250GeV)$. The laser field strength and the laser frequency are chosen as $\epsilon_{0}=10^{7}V.cm^{-1}$ and $\omega=2\, eV$, respectively.}
  \label{fig5}
  \end{figure}
Figures \ref{fig3}, \ref{fig4} and \ref{fig5} show the total cross section versus the net photon number transferred between the colliding system ($e^{+}e^{-}$) and the circularly polarized laser field. The strength and frequency of the laser field in Fig.\ref{fig3}, Fig.\ref{fig4} and Fig.\ref{fig5} are  ($\epsilon_{0}=10^{7}\,\,V.cm^{-1}$ ; $\omega=1.17 eV$), ($\epsilon_{0}=10^{8}\,\,V.cm^{-1}$ ; $\omega=2 eV$) and ($\epsilon_{0}=10^{7}\,\,V.cm^{-1}$ ; $\omega=2 eV$), respectively. The magnitude of the total cross section  varies in a range of a few orders for different photon number $n$. A large number of photons are exchanged between the laser field and the colliding system in all curves and the cutoff number is about $n=\pm 1100$ and $n=\pm 3700$ and $n=\pm400$ in Fig.\ref{fig3}, Fig.\ref{fig4} and Fig.\ref{fig5}, respectively. The process of photon absorption and emission exhibits a symmetric envelope scenario as shown in Fig.\ref{fig3}, Fig.\ref{fig4} and Fig.\ref{fig5}. Comparison between Fig.\ref{fig3} and Fig.\ref{fig5} shows that, for the same laser intensity, the cutoff number decreases as far as the laser frequency increases. Contrarily, for the same value of laser frequency \big(Fig.\ref{fig4} and Fig.\ref{fig5}\big), the required number of exchanged photons to reach the sum-rule increases as far as the laser field strength increases. This symmetry is due to one of the Bessel function properties which is $J_{-n}(z)=(-1)^{n}J_{n}(z)$. The case of net zero-exchanged photons $n=0$ does not correspond to the field-free case, for this situation, the laser field modifies the angular range of the final-state particles momenta and hence the total cross section. Having understood the circularly polarized laser field effect on the Higgs strahlung cross section, we now proceed to explore how the total cross section is influenced when the laser field is embedded in the $e^{+}e^{-}$ collision. In order to get a distinct picture of the pure effects of circularly polarized laser field, we mainly discuss the dependence of the total cross section as a function of the center of mass energy  $\sqrt{s}$ for different number of exchanged photons.
\begin{figure}[H]
  \centering
      \includegraphics[scale=0.8]{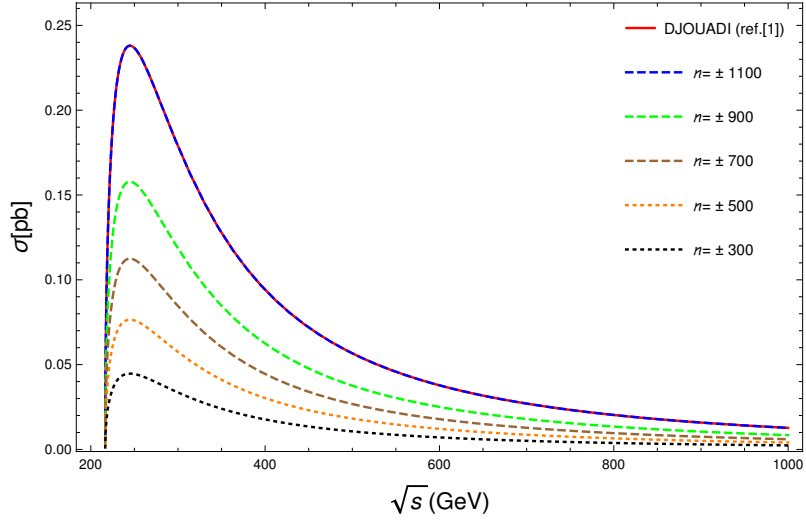}
  \caption{The total cross section of the Higgs strahlung process  $e^{+}e^{-}\rightarrow ZH$ as a function of the center of mass energy $\sqrt{s}$ for different number of exchanged photons.}
  \label{fig6}
\end{figure}
 Figure \ref{fig6} shows the total cross section of the Higgs strahlung production in the absence of an external field and the case of multiphoton process described by equation (\ref{(27)}) where the electron and the positron exchange a number of photons with a circularly polarized laser field. The laser field strength is chosen as $\epsilon_{0}=10^{7}\, V.cm^{-1}$ and its frequency is $\omega=1.17\,eV$. In contrast to the linearly polarized laser field which amplifies the cross section \cite{6}, the circularly polarized laser field decreases the cross section by several order of magnitude by decreasing the transition probability. For a number of exchanged photons $(n=\pm 300)$, the total cross section reaches its maximum at $0.04\,pb$ which is much smaller than the laser-free cross section. As far as the number of exchanged photons increases, the total cross section converges to the laser-free cross section. To reach this convergency, the physical system has to exchange, at least, a number of photons of $(n=\pm 1100)$. This results is in full agreement with the Fig.\ref{fig3} which determines the cutoff number as $(n=\pm 1100)$. So, the multiphoton process cross-section can easily performed by  summing over all possible processes of absorption and emission of the photons. At electron positron collider (CEPC for example), the Higgs bosons are produced through Higgs strahlung and vector fusion process. In the Higgs strahlung process, which is the leading production process at the center of mass energy of $240\,GeV$-$250\,GeV$, the Higgs boson is produced in association with a $Z$-boson. Above the threshold of $ZH$ $(\sqrt{s}=215\,GeV)$, the cross section of Higgs strahlung increases rapidly and reaches its  maximum at $250\,GeV$, then it decreases asymptotically as $(1/s)$ like any typical $s$-channel. Therefore, the vector boson fusion will become the dominant contribution to the Higgs production far beyond the $ZH$ threshold $(\sqrt{s}\geq 500\, GeV)$ \cite{22,23}.\\
 The quantity that measures the ability of a particle collider to produce the required number of events (useful interactions) is called the luminosity and it is defined as the proportionality factor between the number of events per second $dR/dt$ and the total cross section such as: $dR/dt=L\times \sigma$ where $L$ and $\sigma$ are the luminosity of the collider and the total cross section, respectively. The proposed high energy electron positron collider CEPC anticipates integrated luminosity of $3\times 10^{34}cm^{-2}s^{-1}$, $32\times 10^{34}cm^{-2}s^{-1}$ and $10\times 10^{34}cm^{-2}s^{-1}$ for center of mass energy $240\,GeV$, $91\,GeV$ and $160\,GeV$, respectively. As we see from Fig.\ref{fig6}, the cross section diminishes as a function of the photon exchanged number which means that the number of events (Higgs production) rises as far as the number of exchanged photon between the laser field and the colliding system increases. Thus, the expected number of events per second is $0.729\times 10^{-2}$, $0.471\times 10^{-2}$ and $0.14\times 10^{-2}$ for the number of exchanged photons $n=\pm 1100$ (which corresponds to laser-free process), $n=\pm 900$ and $n=\pm 300$, respectively.
 \begin{figure}[H]
  \centering
      \includegraphics[scale=0.8]{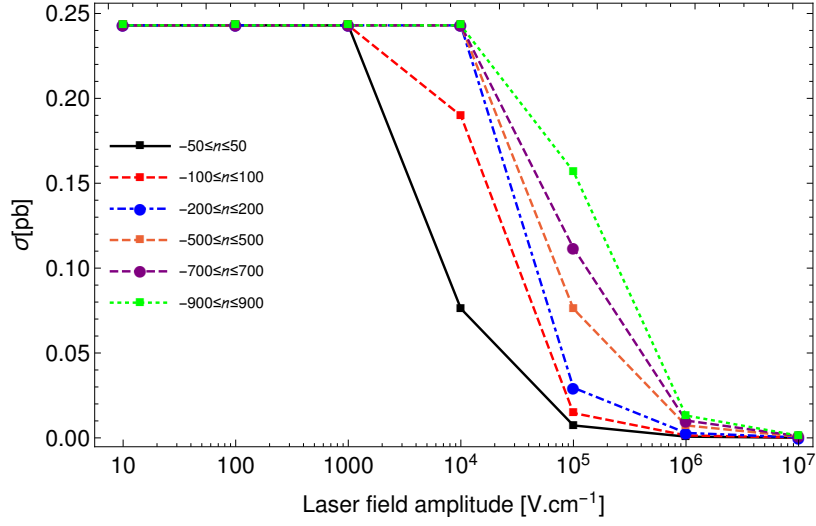}
  \caption{Higgs strahlung process  $e^{+}e^{-}\rightarrow ZH$ cross section as a function of the laser field strength for different exchanged photons number. The center of mass energy and the laser frequency  are taken as $\sqrt{s}=250GeV$ and $\,\omega=0.117 eV$, respectively}
  \label{fig7}
  \end{figure}

   \begin{figure}[H]
  \centering
      \includegraphics[scale=0.8]{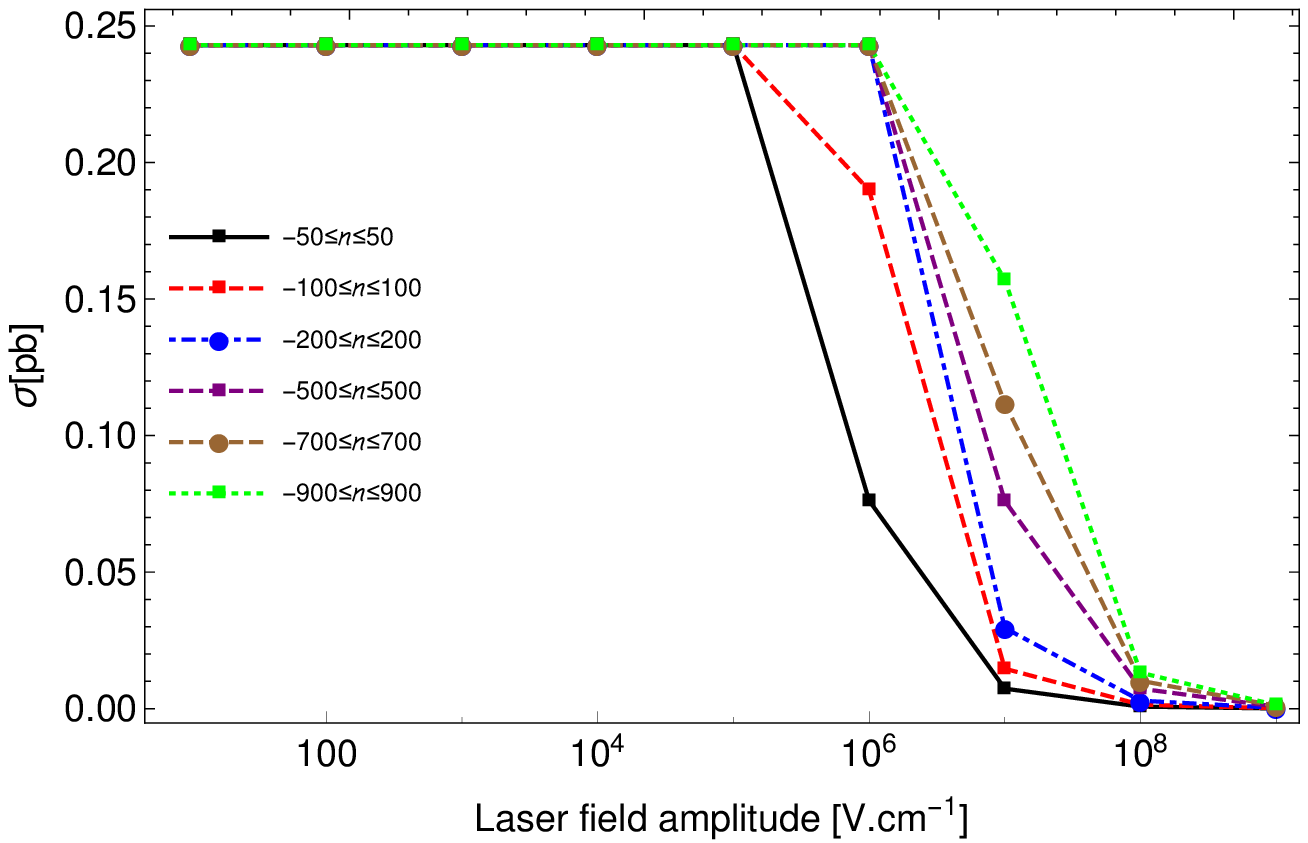}
  \caption{Higgs strahlung process  $e^{+}e^{-}\rightarrow ZH$ cross section as a function of the laser field strength for different exchanged photons number. The center of mass energy and the laser frequency  are taken as $\sqrt{s}=250GeV$ and $\omega=1,17 eV$, respectively.}
  \label{fig8}
  \end{figure}
   \begin{figure}[H]
  \centering
      \includegraphics[scale=0.8]{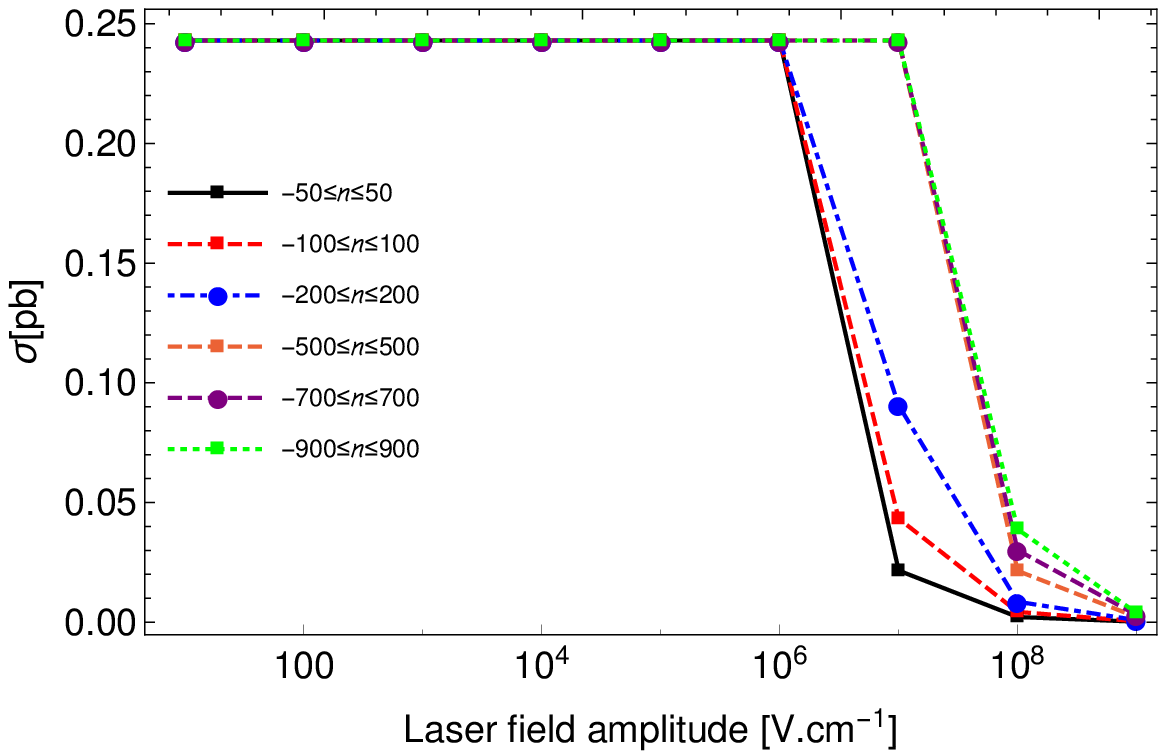}
  \caption{Higgs strahlung process  $e^{+}e^{-}\rightarrow ZH$ cross section as a function of the laser field strength for different exchanged photons number. The center of mass energy and the laser frequency  are taken as $\sqrt{s}=250GeV$ $\omega=2 eV$, respectively.}
  \label{fig9}
  \end{figure}
Figures \ref{fig7}, \ref{fig8} and \ref{fig9} displays the  behavior of the total cross section of the Higgs strahlung production in the center of mass frame  as a function of the circularly polarized laser field strength for different numbers of exchanged  photons and for different laser frequencies. In the range of  small intensities, the laser field has no effect at all on the total cross section, regardless of the laser frequency and the number of exchanged photons. In general, this range depends on the laser field frequency. In Fig.\ref{fig7}, the laser frequency is $\omega=0.117 \,eV$ and the effect of laser begins at $\epsilon_{0}=10^{3}\,V.cm^{-1}$ whereas in Fig.\ref{fig8} and Fig.\ref{fig9} the effect of the laser field begins at $\epsilon_{0}=10^{5}\,V.cm^{-1}$ and $\epsilon_{0}=10^{6}\,V.cm^{-1}$, respectively, for the number of exchanged photons $n=\pm\, 50$. The total cross section diminishes considerably as far as the laser field strength increases, moreover, this decrease depends on the number of exchanged photons and it is due to Bessel functions. For example, the first term $(\sum_{n=-50}^{50}J_{n}^{2}(z)=1)$ in equation (\ref{24}) is equal to $1$ for the laser field strengths between $10$ and $10^{5}\,V.cm^{-1}$, then it begins to decrease and it is equal to $0.313439$ for $\epsilon_{0}=10^{6}\,V.cm^{-1}$. It is obvious that if we sum over the cutoff number of each laser strength and frequency, the curves will not show any dependence of the total cross section on the laser field. \\
At electron positron collider, one of the most used method to measure the Higgs boson is through tagging the Higgs decay final states. This method  allows the number of Higgs events to be counted in several final states. Thus, it allows us to measure the  product of Higgs boson cross section and its decay branching ratio to a given final states $\sigma(e^{+}e^{-}\rightarrow ZH)\times BR(H\rightarrow X \bar{X})$, where $X$ is the possible finale state and  $\sigma(ZH)$ is the cross section of the Higgs strahlung process. Another method is through the recoil mass to the associated $Z$ boson, especially if the $Z$ boson decays to pair of electrons or muons. This recoil mass allow us to better determine the Higgs cross section $\sigma(e^{+}e^{-}\rightarrow ZH)$ and the Higgs coupling $g(HZZ)$ in a model-independent manner \cite{24}. Thus, we have: $\sigma(e^{+}e^{-}\rightarrow ZH)\propto g_{HZZ}^{2}$, $\sigma(e^{+}e^{-}\rightarrow ZH)\times BR(H\rightarrow X \bar{X})\propto g_{HZZ}^{2}g_{HXX}^{2}/\Gamma_{H}$ and  $\sigma(e^{+}e^{-}\rightarrow H\nu_{e}\bar{\nu_{e}})\times BR(H\rightarrow X \bar{X})\propto g_{HZZ}^{2}g_{HWW}^{2}/\Gamma_{H}$. With the recoil method, we can detect the $125\,GeV$ Higgs boson without using its decay informations. In other words, the clean environment and low cross section for background processes at $e^{+}e^{-}$ collider, allow the Higgs strahlung events to be tagged based only on the measurement of the $4$-vector momentum of the $Z$ boson regardless of the Higgs final states in the Higgs decay process. So, we  assume  a  common  modification  to  the Higgs coupling $HZZ$ and $HWW$ for that  the circularly polarized laser field reduces the Higgs strahlung cross section.
\begin{figure}[H]
  \centering
      \includegraphics[scale=0.8]{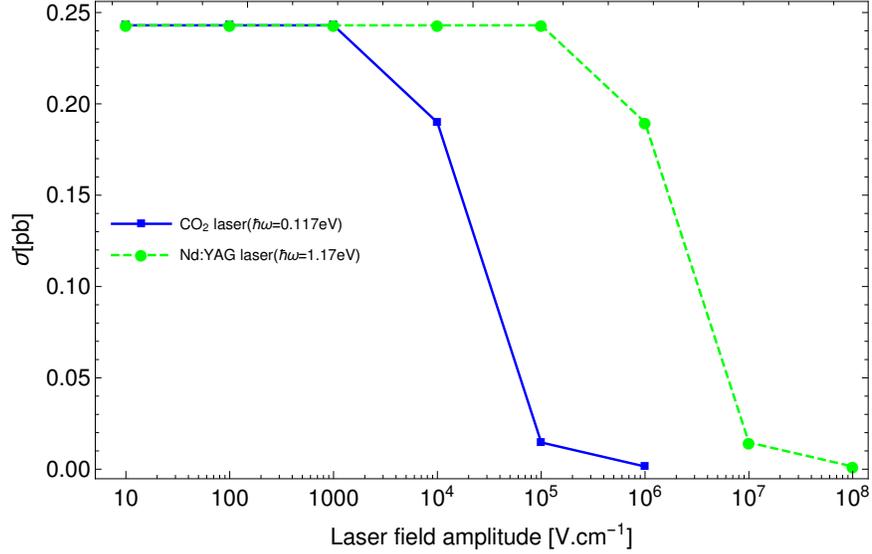}
  \caption{Variation of the total cross section of the Higgs strahlung process  $e^{+}e^{-}\rightarrow ZH$ as a function of the laser field strength $(\epsilon_{0})$ for a Nd:YAG laser $(\omega= 1.17 eV)$ and a CO2 laser$(\omega= 0.117 eV)$. The center of mass energy and the number of exchanged photons are taken as $(-100\leq n\leq 100)$ and $(\sqrt{s}=250GeV)$, respectively.}
  \label{fig10}
\end{figure}
Figure \ref{fig10} illustrates the effect of laser frequency on Higgs strahlung production where the colliding electron and positron exchanges a number of photons  $(-100\leq n\leq 100)$. We consider  two different laser frequencies which are the Nd:YAG laser $(\omega= 1.17 eV)$ and the CO2 laser $(\omega = 0.117 eV)$. Regardless of the laser frequency, each laser field has a precise range of laser field strengths in which no effect on the total cross section is observed. These ranges are $0\leq\epsilon_{0}\leq 10^{3}\,\,V/cm$  and $0\leq\epsilon_{0}\leq 10^{5}\,\,V/cm$ for the CO2 laser and  the Nd:YAG laser, respectively. According to Fig.\ref{fig10}, as far as the laser frequency increases, the effect of the laser field on the total cross section diminishes. Apparently, the total cross section is reduced by the same order of magnitude by which the field strength increases.
\begin{figure}[H]
  \centering
      \includegraphics[scale=0.8]{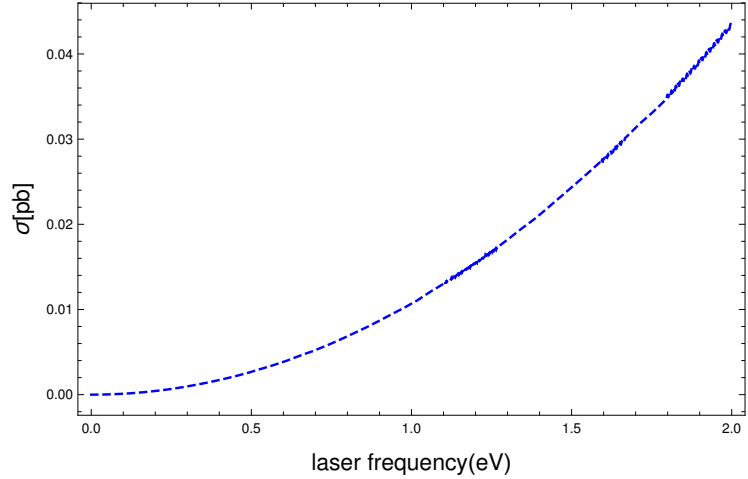}
  \caption{dependence of the Higgs strahlung process's total cross section  as a function of the laser field frequency for $(\epsilon_{0}=10^{7}V/cm)$. The number of exchanged photons and the center of mass energy are taken as  $(-100\leq n\leq 100)$ and $(\sqrt{s}=250GeV)$, respectively.}
  \label{fig11}
\end{figure}
Figure \ref{fig11} shows the variation of the Higgs starhlung multiphon process cross section as a function of the laser frequency. The center of mass energy and the laser field strength are chosen as $(\sqrt{s}=250GeV)$ and $\epsilon_{0}=10^{7}V/cm$, respectively. We have chosen a small number of exchanged photons ($(-100\leq n\leq 100)$) to avoid intensive computing calculations that take a long time. According to Fig.\ref{fig11}, it is clear that at low frequencies, the laser beam does not affects the total cross section. In addition, as the laser frequency overcomes the threshold $(\omega=0.2\, eV)$, the total cross section rises exponentially as much as the laser frequency increases.\\
The Higgs width measurement is an essential ingredient to determine its partial width and coupling constant. The natural width of a $125\,GeV$ Higgs boson, as predicted in the standard model, is only $4.2\,MeV$. As the best precision achieved with current LHC data is $10\,MeV$, the Higgs boson width is very small to be measured experimentally at LHC. The introduction of the laser field in the future  $e^{+} e^{-}$ high luminosity collider could have a significant impact on the measurements.
\section{Conclusion}\label{sec:conclusion}
In the present paper, we investigated, in the framework of the electroweak interaction, the laser-assisted Higgs stahlung process. We showed how the total cross section of the Higgs strahlung production is reduced by the circularly polarized laser field by several order of magnitude. The reduction on the cross section depends on the centre of mass energy of the colliding particles and on the laser's parameters such as the electric field strength  $(\epsilon_{0})$ and its frequency $(\omega)$. The dependence of the total cross section of the multiphoton Higgs strahlung process on the number of exchanged photons is presented in details. Moreover, The variation of the total cross section as a function of the electric field amplitude is shown for the Nd:YAG laser $(\omega= 1.17 eV)$ and the CO2 laser $(\omega = 0.117 eV)$.
\section{Appendix}\label{sec:appedix}
In this appendix, we give the remaining coefficients of the equation (\ref{24}):
\begin{eqnarray}
B&=&\nonumber -\dfrac{e^2}{2((k.p_{1}) (k.p_{2}) M_{Z}^{2})}
  \Big[(2 (k.p_{1}) (k.p_{2}) (2 (a_{1}.p_{3})^2 (g_{a}^{2} + g_{v}^{2}) (k.p_{1}) (k.p_{2}) +2 (a_{2}.p_{3})^2 (g_{a}^{2} + g_{v}^{2})\\&\times &\nonumber (k.p_{1}) (k.p_{2})-2 (a_{1}.p_{3}) (g_{a}^{2} + g_{v}^{2}) ((a_{1}.p_{2}) (k.p_{1}) + (a_{1}.p_{1}) (k.p_{2})) (k.p_{3})+2 (a_{1}.p_{1}) (a_{1}.p_{2}) g_{a}^{2} \\&\times &\nonumber(k.p_{3})^2 + 2 (a_{1}.p_{1}) (a_{1}.p_{2}) g_{v}^{2} (k.p_{3})^2+2 a^{2} g_{a}^{2} (k.p_{3})^2 m_{e}^{2} - 2 a^{2} g_{v}^{2} (k.p_{3})^2 m_{e}^{2} + a^{2} g_{a}^{2} (k.p_{1})^2 M_{Z}^{2} \\&+ &\nonumber a^{2} g_{v}^{2} (k.p_{1})^2 M_{Z}^{2}-2 a^{2} g_{a}^{2} (k.p_{1}) (k.p_{2}) M_{Z}^{2} - 2 a^{2} g_{v}^{2} (k.p_{1}) (k.p_{2}) M_{Z}^{2}+a^{2} g_{a}^{2} (k.p_{2})^2 M_{Z}^{2} + a^{2} g_{v}^{2} \\ &\times &\nonumber(k.p_{2})^2 M_{Z}^{2} -2 a^{2} g_{a}^{2} (k.p_{3})^2 (p_{1}.p_{2}) - 2 a^{2} g_{v}^{2} (k.p_{3})^2 (p_{1}.p_{2})+2 a^{2} g_{a}^{2} (k.p_{1}) (k.p_{3}) (p_{1}.p_{3}) + 2 a^{2}\\&\times &\nonumber g_{v}^{2} (k.p_{1}) (k.p_{3}) (p_{1}.p_{3})+2 a^{2} g_{a}^{2} (k.p_{2}) (k.p_{3}) (p_{1}.p_{3}) + 2 a^{2} g_{v}^{2} (k.p_{2}) (k.p_{3}) (p_{1}.p_{3})+2 a^{2} (g_{a}^{2} + g_{v}^{2}) \\&\times &\nonumber((k.p_{1}) + (k.p_{2})) (k.p_{3}) (p_{2}.p_{3}))-g_{a} g_{v} (k.p_{2}) (-2 (3 (k.p_{1}) - (k.p_{2})) ((k.p_{1}) + (k.p_{2})) M_{Z}^{2} \\&+ & ((k.p_{1})+3 (k.p_{2})) (k.p_{3}) (p_{2}.p_{3})) \epsilon(a_{1},a_{2},k,p_{1})+g_{a} g_{v} (k.p_{1}) (2 ((k.p_{1}) - 3 (k.p_{2})) ((k.p_{1}) \\&+ &\nonumber (k.p_{2})) M_{Z}^{2} + (3 (k.p_{1}) + (k.p_{2})) (k.p_{3}) (p_{1}.p_{3})) \epsilon(a_{1},a_{2},k,p_{2})+g_{a} g_{v} (((k.p_{1}) - (k.p_{2})) ((k.p_{1}) \\&\times &\nonumber(k.p_{2})) (k.p_{3}) (p_{1}.p_{2}) \epsilon(a_{1},a_{2},k,p_{3})+3 ((k.p_{1}) - (k.p_{2}))^2 (k.p_{3})^2 \epsilon(a_{1},a_{2},p_{1},p_{2})+(k.p_{1})^2\\&\times &\nonumber (k.p_{2}) (k.p_{3}) \epsilon(a_{1},a_{2},p_{1},p_{3})+5 (k.p_{1}) (k.p_{2})^2 (k.p_{3}) \epsilon(a_{1},a_{2},p_{1},p_{3})-2 (k.p_{2})^3 (k.p_{3}) \\&\times &\nonumber\epsilon(a_{1},a_{2},p_{1},p_{3}) 2 (k.p_{1})^3 (k.p_{3}) \epsilon(a_{1},a_{2},p_{2},p_{3})-5 (k.p_{1})^2 (k.p_{2}) (k.p_{3})  \epsilon(a_{1},a_{2},p_{2},p_{3})-(k.p_{1})\\&\times &\nonumber (k.p_{2})^2 (k.p_{3})  \epsilon(a_{1},a_{2},p_{2},p_{3})+4 (a_{2}.p_{3}) (k.p_{1}) (k.p_{2}) (k.p_{3})\epsilon(a_{1},k,p_{1},p_{2})-4 (a_{2}.p_{3}) (k.p_{1})^2\\&\times &\nonumber (k.p_{2}) \epsilon(a_{1},k,p_{1},p_{3}) 4 (a_{2}.p_{3}) (k.p_{1}) (k.p_{2})^2 \epsilon(a_{1},k,p_{1},p_{3})+4 (a_{2}.p_{3}) (k.p_{1})^2 (k.p_{2}) \epsilon(a_{1},k,p_{2},p_{3})\\&+ &\nonumber 4 (a_{2}.p_{3}) (k.p_{1}) (k.p_{2})^2 \epsilon(a_{1},k,p_{2},p_{3})-4 (a_{1}.p_{3}) (k.p_{1}) (k.p_{2}) (k.p_{3})\epsilon(a_{2},k,p_{1},p_{2})+4 (a_{1}.p_{3})\\&\times &\nonumber (k.p_{1})^2 (k.p_{2}) \epsilon(a_{2},k,p_{1},p_{3}) 4 (a_{1}.p_{3}) (k.p_{1}) (k.p_{2})^2 \epsilon(a_{2},k,p_{1},p_{3})-(a_{1}.p_{2}) (k.p_{1}) (k.p_{2})\\&\times &\nonumber (k.p_{3}) \epsilon(a_{2},k,p_{1},p_{3})+(a_{1}.p_{2}) (k.p_{2})^2 (k.p_{3})\epsilon(a_{2},k,p_{1},p_{3})-(k.p_{1}) (4 (a_{1}.p_{3}) (k.p_{2}) ((k.p_{1}) \\&+ &\nonumber (k.p_{2})) + (a_{1}.p_{1}) ((k.p_{1})(k.p_{2})) (k.p_{3}))\epsilon(a_{2},k,p_{2},p_{3})))\Big]
\end{eqnarray}
\begin{eqnarray}
C&=&\nonumber -\dfrac{e^2}{2((k.p_{1}) (k.p_{2}) M_{Z}^{2})}\Big[(2 (k.p_{1}) (k.p_{2}) (2 (a_{1}.p_{3})^2 (g_{a}^{2} + g_{v}^{2}) (k.p_{1}) (k.p_{2}) +
 2 (a_{2}.p_{3})^2 (g_{a}^{2} + g_{v}^{2})\\&\times &\nonumber (k.p_{1}) (k.p_{2})-2 (a_{1}.p_{3}) (g_{a}^{2} + g_{v}^{2}) ((a_{1}.p_{2}) (k.p_{1}) + (a_{1}.p_{1}) (k.p_{2})) (k.p_{3})+2 (a_{1}.p_{1}) (a_{1}.p_{2}) g_{a}^{2} \\&\times &\nonumber(k.p_{3})^2 + 2(a_{1}.p_{1}) (a_{1}.p_{2}) g_{v}^{2} (k.p_{3})^2+2 a^{2} g_{a}^{2} (k.p_{3})^2 m_{e}^{2} - 2 a^{2} g_{v}^{2} (k.p_{3})^2 m_{e}^{2}+a^{2} g_{a}^{2} (k.p_{1})^2  M_{Z}^{2} a^{2} \\&\times &\nonumber g_{v}^{2} (k.p_{1})^2 M_{Z}^{2}-2 a^{2} g_{a}^{2} (k.p_{1}) (k.p_{2}) M_{Z}^{2} - 2 a^{2} g_{v}^{2} (k.p_{1}) (k.p_{2}) M_{Z}^{2}+a^{2} g_{a}^{2} (k.p_{2})^2 M_{Z}^{2} + a^{2} g_{v}^{2}\\&\times &\nonumber (k.p_{2})^2  M_{Z}^{2}-2 a^{2} g_{a}^{2} (k.p_{3})^2 (p_{1}.p_{2}) - 2 a^{2} g_{v}^{2} (k.p_{3})^2 (p_{1}.p_{2})+2 a^{2} g_{a}^{2} (k.p_{1}) (k.p_{3}) (p_{1}.p_{3}) + 2 a^{2}\\&\times &\nonumber g_{v}^{2} (k.p_{1})  (k.p_{3}) (p_{1}.p_{3}) + 2 a^{2} g_{a}^{2}(k.p_{2}) (k.p_{3}) (p_{1}.p_{3}) + 2 a^{2} g_{v}^{2} (k.p_{2}) (k.p_{3}) (p_{1}.p_{3})+2 a^{2} (g_{a}^{2} + g_{v}^{2})\\&\times &\nonumber ((k.p_{1})   (k.p_{2})) (k.p_{3}) (p_{2}.p_{3}))+g_{a} g_{v}(k.p_{2}) (-2 (3 (k.p_{1}) - (k.p_{2})) ((k.p_{1}) + (k.p_{2})) M_{Z}^{2} \\&+ &\nonumber((k.p_{1})+ 3 (k.p_{2})) (k.p_{3}) (p_{2}.p_{3})) \epsilon(a_{1},a_{2},k,p_{1})- g_{a} g_{v} (k.p_{1}) (2 ((k.p_{1}) - 3 (k.p_{2})) ((k.p_{1})\\&+ &\nonumber  (k.p_{2})) M_{Z}^{2} + (3 (k.p_{1})(k.p_{2})) (k.p_{3}) (p_{1}.p_{3})) \epsilon(a_{1},a_{2},k,p_{2})+ g_{a} g_{v} ((-(k.p_{1})^2 + (k.p_{2})^2) (k.p_{3})\\&\times &\nonumber (p_{1}.p_{2})\epsilon(a_{1},a_{2},k,p_{3}) 3 ((k.p_{1}) - (k.p_{2}))^2 (k.p_{3})^2 \epsilon(a_{1},a_{2},p_{1},p_{2})-(k.p_{1})^2 (k.p_{2}) (k.p_{3})\\&\times &\nonumber\epsilon(a_{1},a_{2},p_{1},p_{3})- 5 (k.p_{1}) (k.p_{2})^2 (k.p_{3}) \epsilon(a_{1},a_{2},p_{1},p_{3})+2 (k.p_{2})^3 (k.p_{3}) \epsilon(a_{1},a_{2},p_{1},p_{3})- 2 \\&\times &\nonumber(k.p_{1})^3 (k.p_{3}) \epsilon(a_{1},a_{2},p_{2},p_{3})+5 (k.p_{1})^2 (k.p_{2}) (k.p_{3}) \epsilon(a_{1},a_{2},p_{2},p_{3})+(k.p_{1}) (k.p_{2})^2 (k.p_{3}) \\&\times &\nonumber\epsilon(a_{1},a_{2},p_{2},p_{3})- 4 (a_{2}.p_{3}) (k.p_{1}) (k.p_{2}) (k.p_{3}) \epsilon(a_{1},k,p_{1},p_{2})+4 (a_{2}.p_{3}) (k.p_{1})^2 (k.p_{2}) \\&\times &\nonumber\epsilon(a_{1},k,p_{1},p_{3})+4 (a_{2}.p_{3}) (k.p_{1}) (k.p_{2})^2 \epsilon(a_{1},k,p_{1},p_{3})4 (a_{2}.p_{3}) (k.p_{1})^2 (k.p_{2})\epsilon(a_{1},k,p_{2},p_{3})\\&- &\nonumber 4 (a_{2}.p_{3}) (k.p_{1}) (k.p_{2})^2 \epsilon(a_{1},k,p_{2},p_{3})+4 (a_{1}.p_{3}) (k.p_{1}) (k.p_{2}) (k.p_{3}) \epsilon(a_{2},k,p_{1},p_{2})-4 (a_{1}.p_{3})\\&\times &\nonumber (k.p_{1})^2 (k.p_{2}) \epsilon(a_{2},k,p_{1},p_{3})-4 (a_{1}.p_{3}) (k.p_{1}) (k.p_{2})^2 \epsilon(a_{2},k,p_{1},p_{3})+(a_{1}.p_{2}) (k.p_{1}) (k.p_{2}) (k.p_{3}) \\&\times &\nonumber\epsilon(a_{2},k,p_{1},p_{3})-(a_{1}.p_{2}) (k.p_{2})^2 (k.p_{3}) \epsilon(a_{2},k,p_{1},p_{3})(k.p_{1}) (4 (a_{1}.p_{3}) (k.p_{2}) ((k.p_{1}) + (k.p_{2})) \\&+ & (a_{1}.p_{1}) ((k.p_{1}) - (k.p_{2})) (k.p_{3})) \epsilon(a_{2},k,p_{2},p_{3})))\Big]
\end{eqnarray}
\begin{eqnarray}
D&=&\nonumber\dfrac{2\, e}{((k.p_{1}) (k.p_{2}) M_{Z}^{2})}\Big[(g_{a}^{2}+g_{v}^{2}) (a^{2} e^2 (k.p_{3}) ((a_{1}.p_{3}) ((k.p_{1}) - (k.p_{2})) + (-(a_{1}.p_{1})+(a_{1}.p_{2})) \\& \times &\nonumber(k.p_{3}))+(a_{1}.p_{2}) (k.p_{1}) (-((k.p_{1}) + (k.p_{2})) M_{Z}^{2} - 2 (k.p_{3}) (p_{1}.p_{3}))+(k.p_{2}) ((a_{1}.p_{1}) ((k.p_{1}) \\&+&(k.p_{2})) M_{Z}^{2} + 2 (a_{1}.p_{3}) (k.p_{1}) ((p_{1}.p_{3})-(p_{2}.p_{3})) +2 (a_{1}.p_{1}) (k.p_{3}) (p_{2}.p_{3})))\Big]
\end{eqnarray}
\begin{eqnarray}
E&=&\nonumber\dfrac{2\, e}{((k.p_{1}) (k.p_{2}) M_{Z}^{2})}\Big[((g_{a}^{2}+g_{v}^{2}) (k.p_{1}) (k.p_{2}) (a^{2} e^2 (k.p_{3}) ((a_{1}.p_{3}) ((k.p_{1}) - (k.p_{2})) + (-(a_{1}.p_{1})\\&+ &\nonumber(a_{1}.p_{2})) (k.p_{3})) + (a_{1}.p_{2}) (k.p_{1}) (-((k.p_{1}) + (k.p_{2})) M_{Z}^{2} - 2 (k.p_{3}) (p_{1}.p_{3}))+(k.p_{2}) ((a_{1}.p_{1}) \\&\times &\nonumber((k.p_{1}) + (k.p_{2})) M_{Z}^{2} + 2 (a_{1}.p_{3}) (k.p_{1}) ((p_{1}.p_{3}) - (p_{2}.p_{3}))+2 (a_{1}.p_{1}) (k.p_{3}) (p_{2}.p_{3})))+g_{a} g_{v}\\&\times &\nonumber(((k.p_{1}) - (k.p_{2})) (a^{2} e^2 (k.p_{3})^2 - 4 (k.p_{1}) (k.p_{2}) M_{Z}^{2})\epsilon(a_{2},k,p_{1},p_{2})+(k.p_{2}) ((k.p_{1}) + (k.p_{2})) \\&\times &\nonumber(a^{2} e^2 (k.p_{3}) + 2 (k.p_{1}) (p_{2}.p_{3})) \epsilon(a_{2},k,p_{1},p_{3}) +
(k.p_{1}) (((k.p_{1}) + (k.p_{2})) (a^{2} e^2 (k.p_{3}) + 2 (k.p_{2}) \\&\times &\nonumber(p_{1}.p_{3})) \epsilon(a_{2},k,p_{2},p_{3})+2 (k.p_{2}) (-(k.p_{1}) +(k.p_{2})) (-(k.p_{3})\epsilon(a_{2},p_{1},p_{2},p_{3})\\&+ &(a_{2}.p_{3}) \epsilon(k,p_{1},p_{2},p_{3})))))\Big]
\end{eqnarray}
\begin{eqnarray}
F&=&\nonumber - \dfrac{4\, e^2}{2((k.p_{1}) (k.p_{2}) M_{Z}^{2})}\Big[(g_{a}^{2} + g_{v}^{2}) (((a_{1}.p_{3}) - (a_{2}.p_{3})) ((a_{1}.p_{3}) + (a_{2}.p_{3})) (k.p_{1}) \\&\times &(k.p_{2})-(a_{1}.p_{3}) ((a_{1}.p_{2}) (k.p_{1}) + (a_{1}.p_{1}) (k.p_{2})) (k.p_{3}) + (a_{1}.p_{1}) (a_{1}.p_{2})(k.p_{3})^2))\Big]
\end{eqnarray}

\end{document}